\documentclass{blois}

\def\be{\begin{equation}}
\def\ee{\end{equation}}
\def\bea{\begin{eqnarray}}
\def\eea{\end{eqnarray}}

\newcommand{\Photo}{}

\begin{document}
\vspace*{4cm}
\title{Exotic hadron spectroscopy at the LHCb experiment}

\author{ G. A. Cowan, on behalf of the LHCb collaboration }

\address{School of Physics and Astronomy, University of Edinburgh, EH9 3FD, UK}

\maketitle\abstracts{
The LHCb experiment is designed to study the decays and properties of heavy flavoured hadrons
produced in the forward region from proton-proton collisions at the CERN Large Hadron Collider. During
Run 1, it has recorded the world's largest data sample of beauty and charm hadrons, enabling
precise studies into the spectroscopy of such particles, including discoveries of new states and
measurements of their masses, widths and quantum numbers. An overview of
recent LHCb results in the area of exotic hadron spectroscopy is presented, focussing on the
discovery of the first pentaquark states in the $\Lambda_b^0 \to J/\psi p K^-$ channel and a
search for them in the related $\Lambda_b^0 \to J/\psi p\pi^-$ mode. The LHCb
non-confirmation of the D0 tetraquark candidate in the $B_s^0\pi^+$ invariant mass spectrum
is presented.}

\section{Introduction}

Exotic hadrons are defined as hadrons having internal 
structures more complex than the $q\overline{q}$ mesons or $qqq$ baryons systems
in the quark model. Since 2003 when the $X(3872)$ state was
discovered by the Belle collaboration~\cite{Choi:2003ue} the search for and study of
new exotic particles has become a hot topic since, spurred on by a wealth of new data
from many particle collider experiments. This data has allowed the mass, width and
quantum numbers of many states to be measured and new states discovered.
However, the lack of any clear pattern necessitates a new programme of experimental and
theoretical study to understand the strong interaction dynamics that can cause their production and structure.

These proceedings will discuss recent results from the LHCb collaboration in the 
area of exotic hadron spectroscopy, focussing on the recent pentaquark 
observation in the $\Lambda_b^0 \to J/\psi p K^-$ channel and
a new model-independent analysis to further support the case for the
exotic nature of this state. Evidence for exotic structures in the
Cabibbo suppressed $\Lambda_b^0 \to J/\psi p\pi^-$ mode will be presented
as will the non-confirmation of an exotic resonance in the $B_s^0\pi^+$ invariant mass spectrum.
Further information on exotic hadron spectroscopy can be found in Ref.~\cite{stone}.

\section{Pentaquark observation}\label{sec:pentaquark}

\begin{figure}[t]
\includegraphics[scale=0.40]{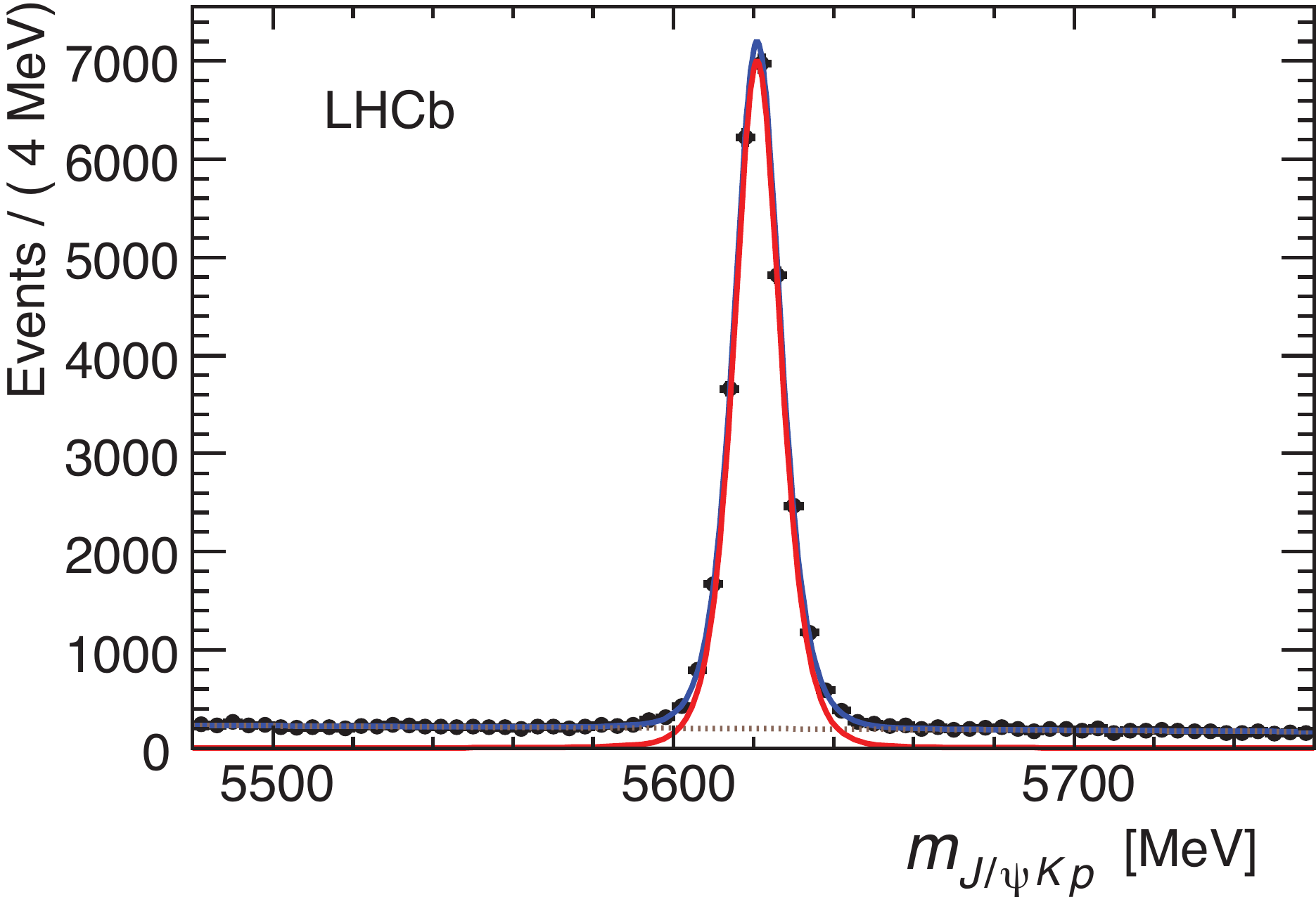}
\includegraphics[scale=0.35]{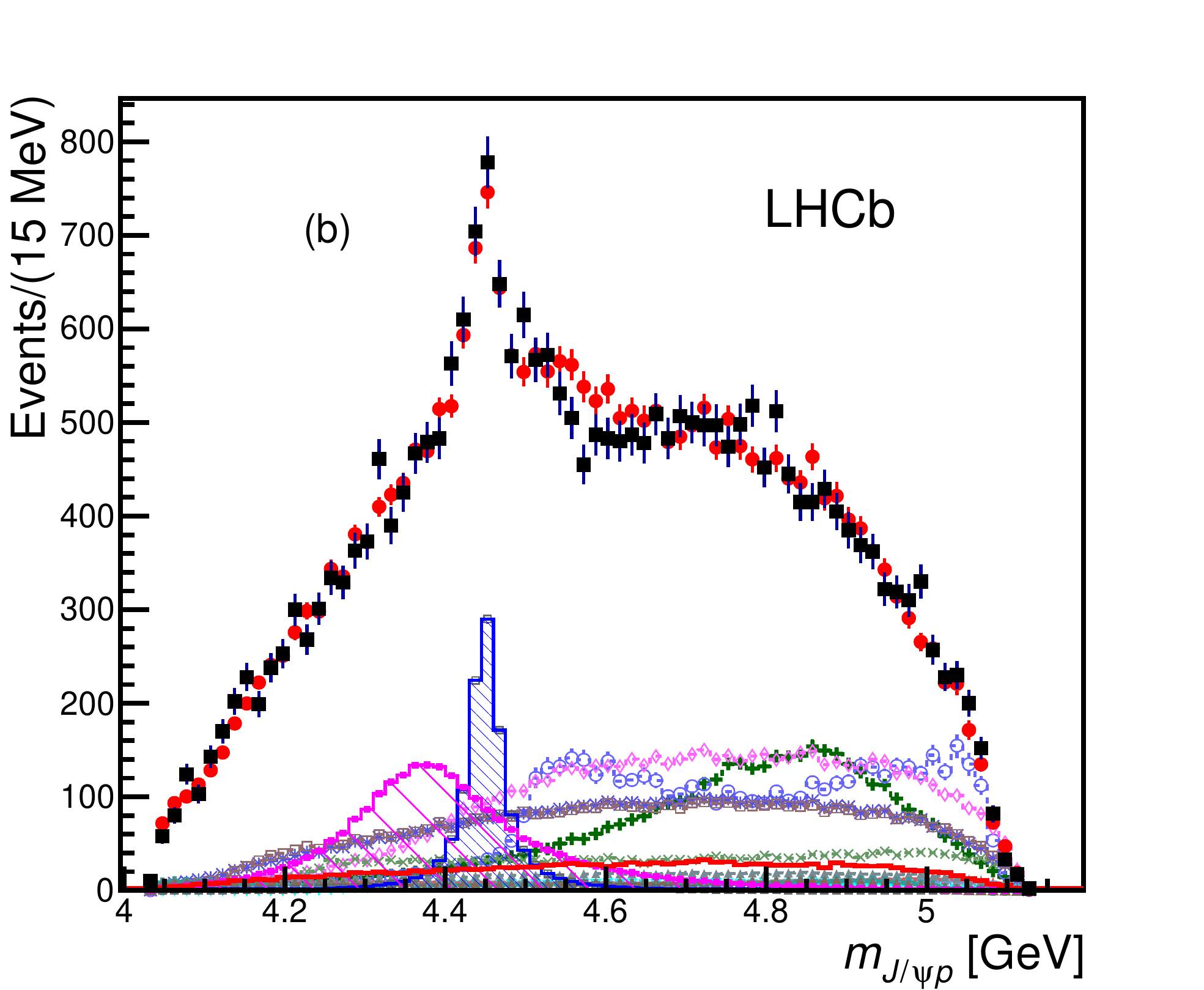}
\put(-370,90){(a)}
\caption{(a) Distribution of $m(J/\psi p K)$ for selected  $\Lambda_b^0 \to J/\psi p K^-$ candidates (black points) and the fit projection.
(b) Fit projections for $m(J/\psi p)$ for the reduced $\Lambda^*$ model with two $P^+_c$ states. The data are shown as solid (black) squares, while the solid (red) points show the fit. The solid (red) histogram shows the background distribution. The (blue) open squares with the shaded histogram represent the $P_c(4450)^+$ state and the shaded histogram topped with (purple) filled squares represents the $P_c(4380)^+$
state. Each $\Lambda^*$ component is also shown.}
\label{fig:JpsipK}
\end{figure}

Figure~\ref{fig:JpsipK} (a) shows the distribution of $m(J/\psi p K^-)$ from the
selected $\Lambda_b^0 \to J/\psi p K^-$ candidates~\cite{Aaij:2015tga} in the LHCb Run 1 sample. 
There are approximately $26000$ signal events with a
$5.4\%$ background within a $2\sigma$ window around the known $\Lambda_b^0$ mass.
A six-dimensional amplitude model is constructed to fully describe the structure of
the $\Lambda_b^0 \to J/\psi p K^-$ decay. All known $\Lambda^* \to p K^-$
states and two exotic $P_c^+ \to J/\psi p$ states are included in the model, with their
line shapes being described using relativistic Breit-Wigner functions.
Fully simulated events are used
to correct for the detector and selection efficiencies. The amplitude model 
containing only $\Lambda^*$ resonances is insufficient to describe the data
and two interfering $P_c^+$ states with opposite parity are required (see Figure~\ref{fig:JpsipK}
(b)) both of which are statistically significant at larger than $9\sigma$, evaluated using
pseudoexperiments.
The low mass state ($4380 \pm 8 \pm 29$ MeV) is wide ($205 \pm 18 \pm 86$ MeV) with $J^P=3/2^-$
and a fit fraction of $(8.4\pm 0.7\pm 4.2)\%$
while the high mass state ($4449.8 \pm 1.7 \pm 2.5$ MeV) is narrow ($39 \pm 5 \pm 19$ MeV)
with $J^P=5/2^+$ and a fit fraction of $(4.1\pm 0.5\pm 1.1)\%$.
Other $J^P$ combinations give similar fit qualities. The main systematic
uncertainties in the result come from the limited knowledge of the $\Lambda^*$ spectrum.

\begin{figure}[t]
\includegraphics[scale=0.40]{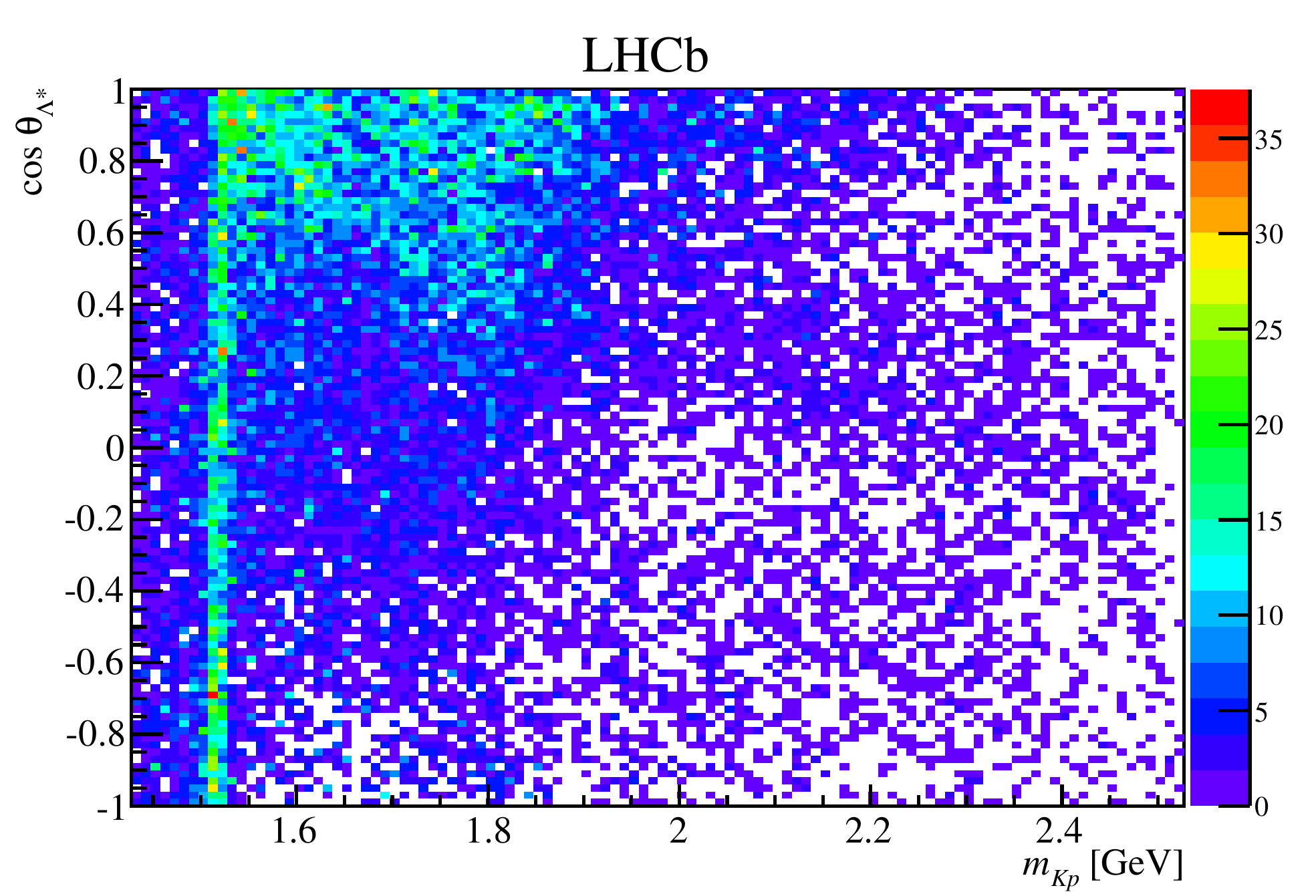}
\includegraphics[scale=0.40]{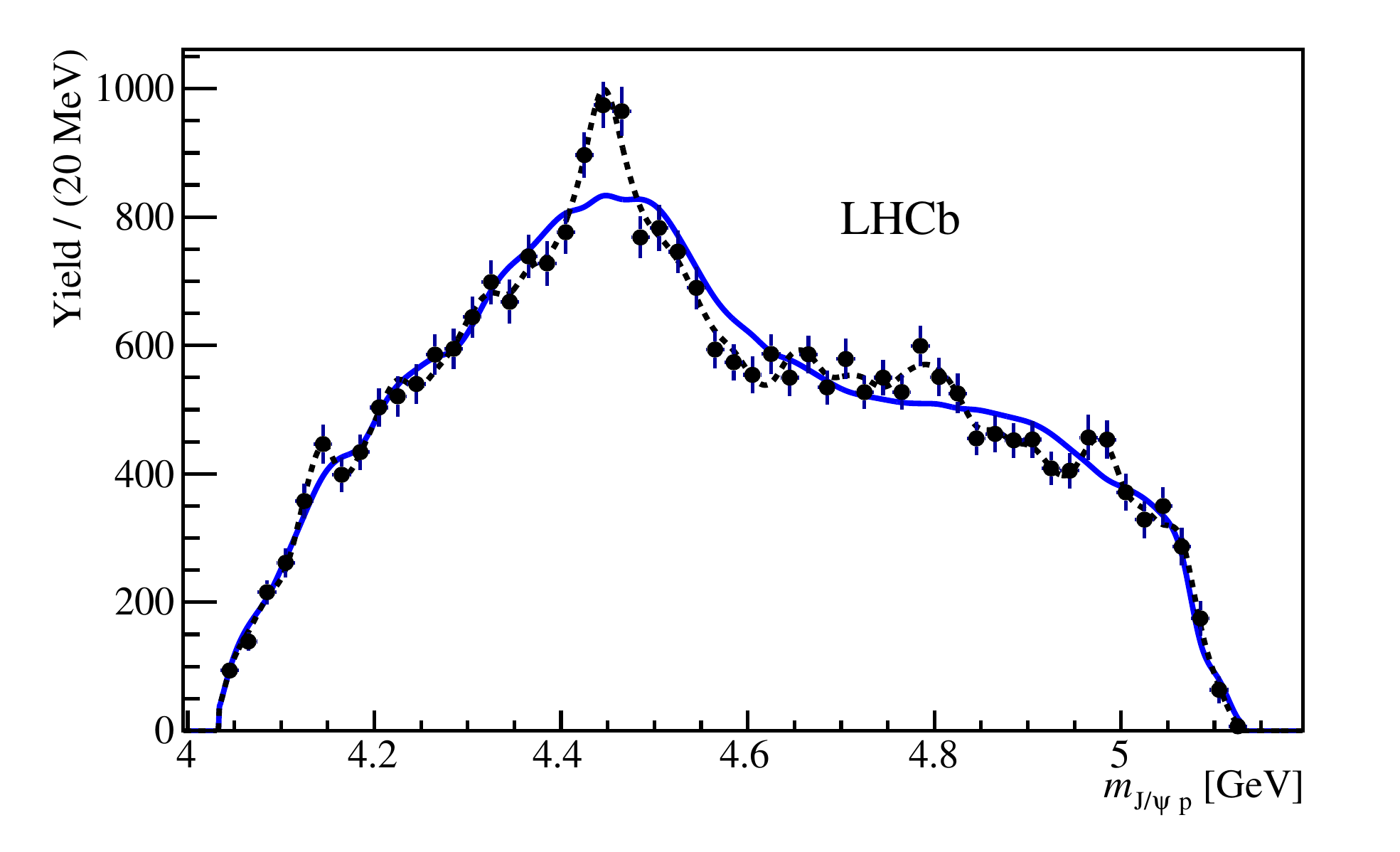}
\put(-380,145){(a)}
\put(-180,90){(b)}
\caption{(a) Background-subtracted and efficiency-corrected distribution of the cosine of the $\Lambda^*$ helicity angle versus $m(Kp)$.
(b) Efficiency-corrected and background-subtracted $m(J/\psi p)$ distribution of the data (black points), with ${\cal F}(m_{J/\psi p}|H_0)$
(solid blue line) and ${\cal F}(m_{J/\psi p}|H_0)$ (dashed black line) superimposed.}
\label{fig:JpsipK_MI}
\end{figure}

The case for the exotic nature of the $P_c^+$ contributions to the $\Lambda_b^0 \to J/\psi p K^-$
channel is strengthened via a model independent analysis in Ref.~\cite{Aaij:2016phn} in which
no assumption is made regarding the number and type of $\Lambda^*$ resonances. Instead, the analysis
simply tries to answer the question if the structure observed in the distribution of $m(J/\psi p K^-)$
can be explained via reflections from the interfering $\Lambda^*$ states that are
visible in $m(p K^-)$. Only
physical arguments are employed to restrict the maximal $\Lambda^*$ spin that can contribute
at a given value of $m(p K^-)$.
Figure~\ref{fig:JpsipK_MI} (a) shows the two-dimensional Dalitz plot of the
$\Lambda_b^0 \to J/\psi p K^-$ data, which is decomposed into a set of Legendre polynomials. 
These polynomials are then used to weight the phase-space simulated $\Lambda_b^0 \to J/\psi p K^-$
sample to see if it can reproduce the $m(J/\psi p K^-)$ distribution in Figure~\ref{fig:JpsipK_MI} (b).
Pseudoexperiments are  generated to build a statistic that can be used to test the null hypothesis that
only $\Lambda^*$ resonances are required. These find that in the data, the null hypothesis is ruled out at
more than $9\sigma$. 

\section{Evidence for exotic hadron contributions to $\Lambda_b^0 \to J/\psi p\pi^-$ decays}

\begin{figure}[t]
\includegraphics[scale=0.40]{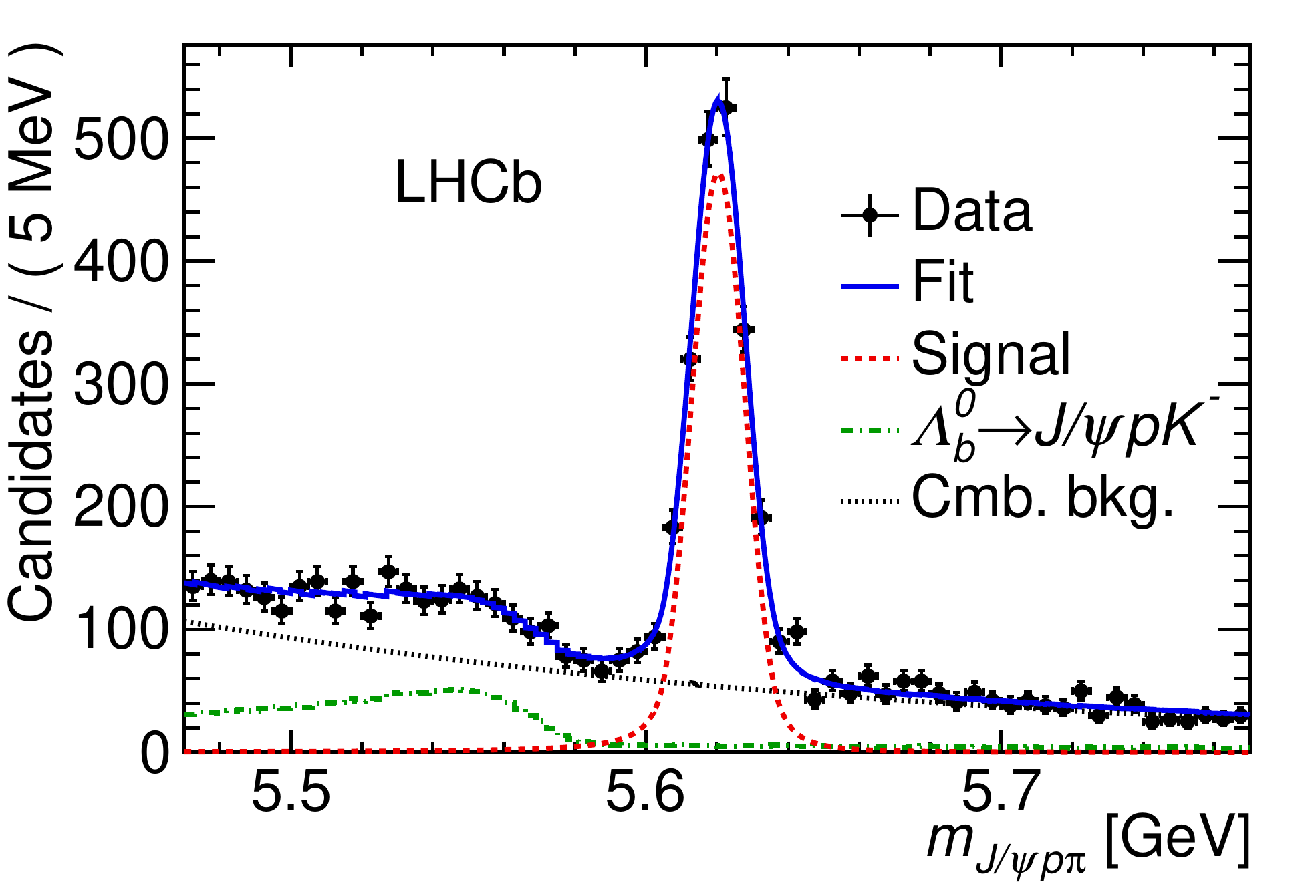}
\includegraphics[scale=0.35]{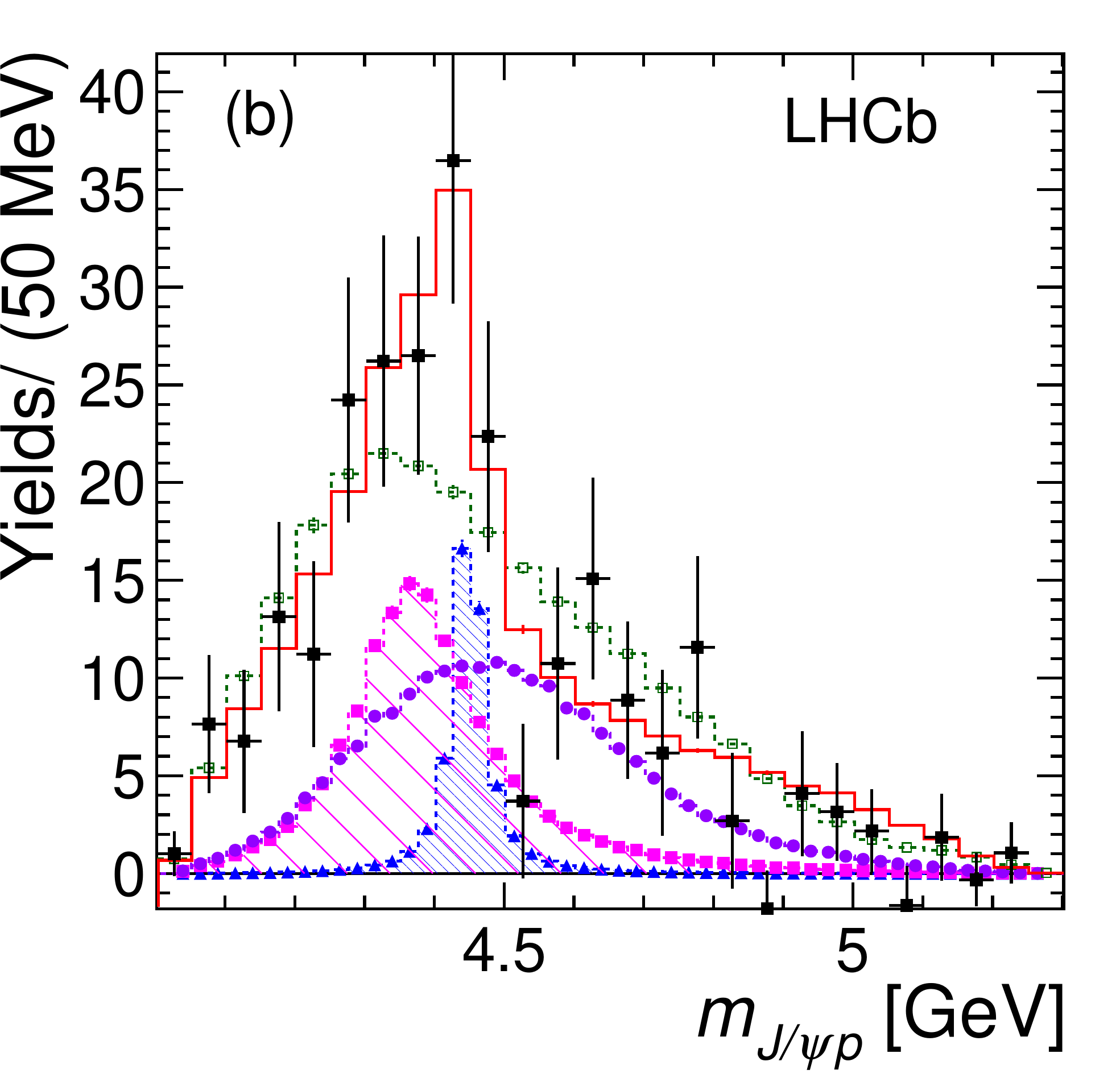}
\put(-380,90){(a)}
\caption{(a) Distribution of $m(J/\psi p\pi^-)$ for selected $\Lambda_b^0 \to J/\psi p\pi^-$ candidates (points) and the fit projection.
(b) Background-subtracted
data (black points) and total fit (red line) projections onto $m(J/\psi p)$ for the region with $m(p\pi^-) >1.8$ GeV.
The $N^*$-only model is shown by the green histogram, which is clearly not a good fit to the data. The
$P_c(4380)^+$, $P_c(4450)^+$ and $Z_c(4200)^-$ components are shown by the pink, blue and purple histograms, respectively.}
\label{fig:Jpsippi}
\end{figure}

Following the observation of the $P_c^+$ states described above, it is important to search for other
production and decay modes in order to help understand if they are genuinely exotic baryons or
perhaps some sort of kinematical effect~\cite{Wang:2015pcn}.
In Ref.~\cite{Aaij:2016ymb} the LHCb collaboration has observed the
Cabibbo suppressed $\Lambda_b^0 \to J/\psi p\pi^-$ 
decay and measured its branching ratio to the Cabibbo favoured mode to be approximately $8\%$. 
As described above, an amplitude model is constructed to describe the three mutually 
interfering decay chains ($\Lambda_b^0 \to J/\psi N^* (\to p\pi^-)$, 
$\Lambda_b^0 \to P_c^+ (\to J/\psi p) \pi^-$ and
$\Lambda_b^0 \to Z_c^- (\to J/\psi \pi^-) p$), which is then fit to the approximately 1900 signal
events that are isolated using the $m(J/\psi p\pi^-)$ distribution shown in Figure~\ref{fig:Jpsippi} (a). 
The fit model including only the well-established $N^*$ states is insufficient to describe the 
data and show significant improvement when exotic contributions are included. When all three
exotics are included (taking their parameters from Refs.~\cite{Aaij:2015tga}
and~\cite{Chilikin:2014bkk}) there is $3.1\sigma$ evidence, with their fit fractions measured
to be $(5.1 \pm 1.5^{+2.1}_{-1.6})\%\ (P_c(4380)^+)$, $(1.6^{+0.8 +0.6}_{-0.6 - 0.5})\%\ (P_c(4450)^+)$
and $(7.7 \pm 2.8^{+3.4}_{-4.0})\%\ (Z_c(4200)^-)$, which are consistent with those
in Ref.~\cite{Aaij:2015tga}, accounting for the Cabibbo suppression. No single 
$P_c^+$ or $Z_c^-$ component makes a significant difference to the model. 
Figure~\ref{fig:Jpsippi} (b) shows the distribution of $m(J/\psi p)$ for the region of high $m(p\pi^-)$
along with the fit projection. The main systematic uncertainties come from the the knowledge of
the exotic masses, widths, $J^P$ and the knowledge about the content of the $N^*$ spectrum.

\section{Search for structure in the $B_s^0\pi^+$ invariant mass spectrum}

\begin{figure}[t]
\includegraphics[scale=0.40]{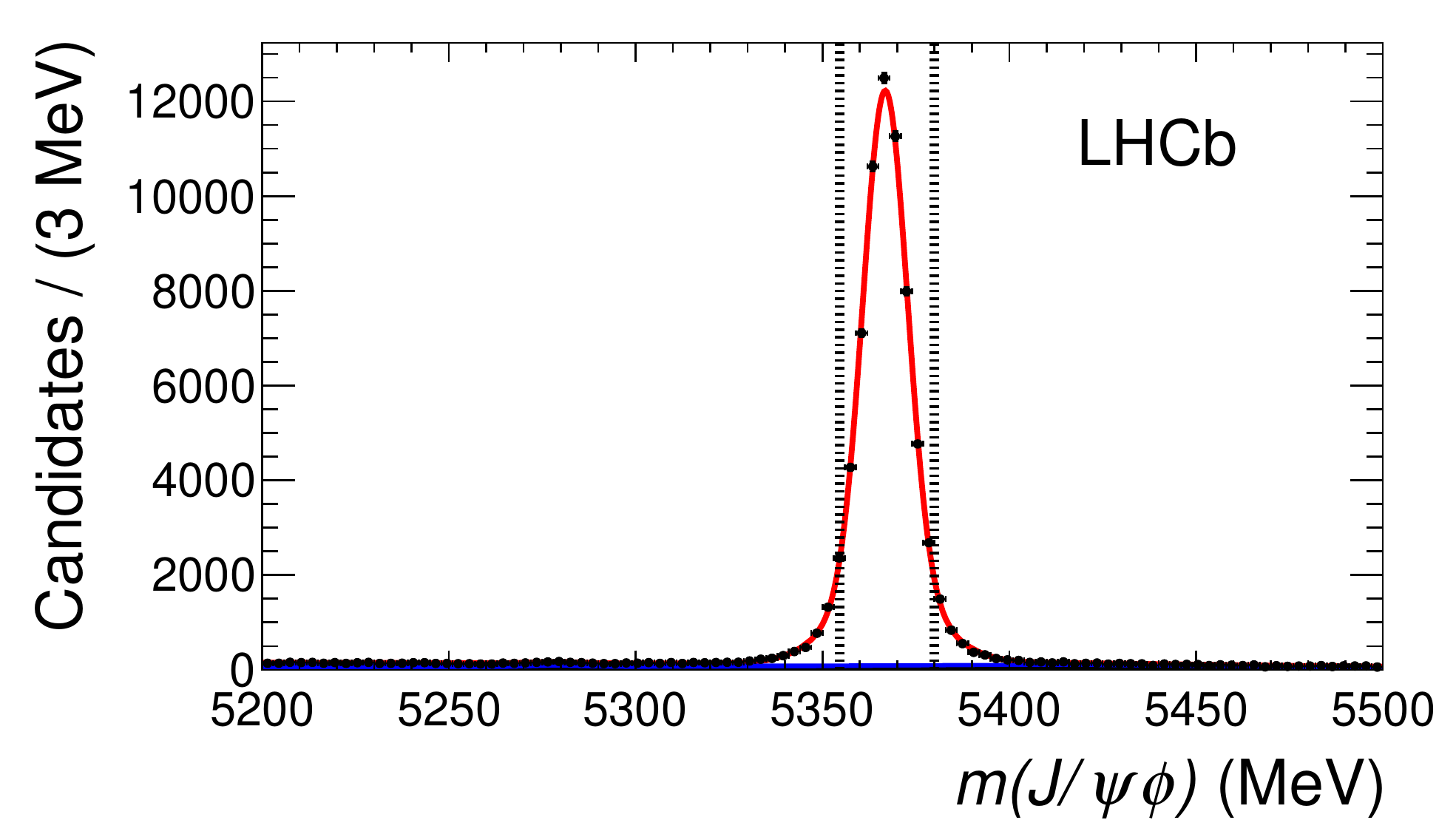}
\includegraphics[scale=0.3725]{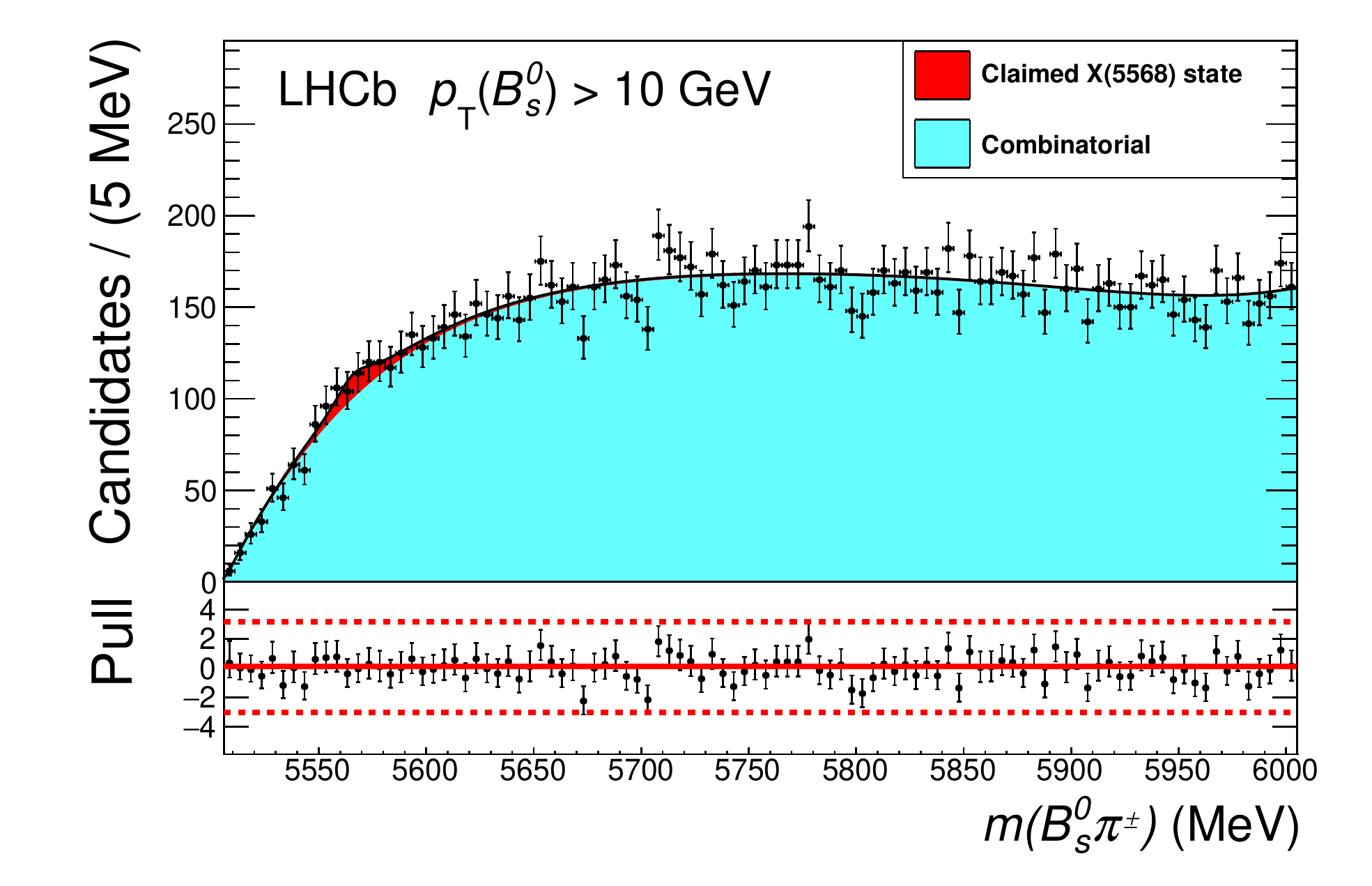}
\put(-390,100){(a)}
\put(-160,100){(b)}
\caption{(a) Selected $B_s^0 \to J/\psi \phi$ decays with $p_{\rm T}(B_s^0)>5$ GeV, where the $B_0^s$ signal window requirements are indicated by dotted lines. (b) Results of the fit to the $B_s^0\pi^\pm$ mass distribution for all candidates with $p_{\rm T}(B_s^0)>10$ GeV.
The component for the claimed $X(5568)$ state is included in the fit but is not significant.}
\label{fig:x5568}
\end{figure}

In Ref.\cite{D0:2016mwd} the D0 collaboration claimed evidence for a new state, the $X(5568)$,
after finding an enhancement near threshold in the $B_s^0\pi^+$ invariant mass spectrum and
measuring the production fraction of this state to be $\rho = (8.6\pm1.9\pm1.4)\%$ of the $B_s^0$
meson production.
Using a dataset twenty times larger than D0's, the LHCb collaboration performed
a  search~\cite{Aaij:2016iev} for the $X(5568)$ using very clean samples of $B_s^0\to D_s^-\pi^+$
and $B_s^0\to J/\psi\phi$ decays (see Figure~\ref{fig:x5568}(a)). Inspection of the
$B_s^0\pi^+$ mass distribution in various kinematical regimes did not show any evidence of an
enhancement, allowing limits to be set on its production as $\rho (p_T(B_s^0) > 10$ GeV $) < 0.021 (0.024)$
at 90 (95)\% confidence level. Limits were also placed as a function of the supposed mass and width of the 
$X$ state.

\vspace{-0.2cm}
\section{Summary}

Recent results from the LHCb collaboration in the area of exotic hadron spectroscopy have been presented, 
showing more detailed studies of the pentaquark states discovered last year and evidence for their presence
in a new decay mode. In addition, the LHCb collaboration has not confirmed the near-threshold enhancement 
in the $B_s^0\pi^+$ invariant mass spectrum, which was reported by the D0 collaboration.

\vspace{-0.2cm}
\section*{Acknowledgements}
\vspace{-0.2cm}

I would like to thank the organisers of Rencontres de Blois for creating a very enjoyable conference atmosphere.
I acknowledge support from STFC grant ST/K004646/1.


\vspace{-0.2cm}
\section*{References}
\vspace{-0.2cm}

\begin{thebibliography}{99}

\bibitem{Choi:2003ue}
  S.~K.~Choi {\it et al.} [Belle Collaboration],
  Phys.\ Rev.\ Lett.\  {\bf 91} (2003) 262001
  doi:10.1103/PhysRevLett.91.262001
  [hep-ex/0309032].
  
  \bibitem{stone} S.~Stone, these proceedings.

  

%
%



\bibitem{Aaij:2015tga}
  R.~Aaij {\it et al.} [LHCb Collaboration],
  Phys.\ Rev.\ Lett.\  {\bf 115} (2015) 072001
  doi:10.1103/PhysRevLett.115.072001
  [arXiv:1507.03414 [hep-ex]].

\bibitem{Aaij:2016phn}
  R.~Aaij {\it et al.} [LHCb Collaboration],
  Phys.\ Rev.\ Lett.\  {\bf 117} (2016) no.8,  082002
  doi:10.1103/PhysRevLett.117.082002
  [arXiv:1604.05708 [hep-ex]].

\bibitem{Aubert:2008aa}
  B.~Aubert {\it et al.} [BaBar Collaboration],
  Phys.\ Rev.\ D {\bf 79} (2009) 112001
  doi:10.1103/PhysRevD.79.112001
  [arXiv:0811.0564 [hep-ex]].

\bibitem{Aaij:2014zoa}
  R.~Aaij {\it et al.} [LHCb Collaboration],
  JHEP {\bf 1407} (2014) 103
  doi:10.1007/JHEP07(2014)103
  [arXiv:1406.0755 [hep-ex]].

\bibitem{Wang:2015pcn}
  E.~Wang, H.~X.~Chen, L.~S.~Geng, D.~M.~Li and E.~Oset,
  Phys.\ Rev.\ D {\bf 93} (2016) no.9,  094001
  doi:10.1103/PhysRevD.93.094001
  [arXiv:1512.01959 [hep-ph]].

\bibitem{Cheng:2015cca}
  H.~Y.~Cheng and C.~K.~Chua,
  Phys.\ Rev.\ D {\bf 92} (2015) no.9,  096009
  doi:10.1103/PhysRevD.92.096009
  [arXiv:1509.03708 [hep-ph]].

\bibitem{Aaij:2016ymb}
  R.~Aaij {\it et al.} [LHCb Collaboration],
  Phys.\ Rev.\ Lett.\  {\bf 117} (2016) no.8,  082003
   Addendum: [Phys.\ Rev.\ Lett.\  {\bf 117} (2016) no.10,  109902]
  doi:10.1103/PhysRevLett.117.082003, 10.1103/PhysRevLett.117.109902
  [arXiv:1606.06999 [hep-ex]].

\bibitem{Chilikin:2014bkk}
  K.~Chilikin {\it et al.} [Belle Collaboration],
  Phys.\ Rev.\ D {\bf 90} (2014) no.11,  112009
  doi:10.1103/PhysRevD.90.112009
  [arXiv:1408.6457 [hep-ex]].

\bibitem{D0:2016mwd}
  V.~M.~Abazov {\it et al.} [D0 Collaboration],
  Phys.\ Rev.\ Lett.\  {\bf 117} (2016) no.2,  022003
  doi:10.1103/PhysRevLett.117.022003
  [arXiv:1602.07588 [hep-ex]].

\bibitem{Aaij:2016iev}
  R.~Aaij {\it et al.} [LHCb Collaboration],
  Phys.\ Rev.\ Lett.\  {\bf 117} (2016) 152003
  doi:10.1103/PhysRevLett.117.152003
  [arXiv:1608.00435 [hep-ex]].
 

\end{thebibliography}
\end{document}